\documentclass[aps, prd, showpacs, superscriptaddress, nofootinbib, twocolumn]{revtex4}

\usepackage{amssymb, amsmath, bm, dcolumn, epsf, graphicx, latexsym, slashed, simplewick}

\def\be{\begin{equation}}
\def\ee{\end{equation}}
\def\bea{\begin{eqnarray}}
\def\eea{\end{eqnarray}}

\begin{document}

\title{Matter creation in a nonsingular bouncing cosmology}

\author{Jerome Quintin}
\email{jquintin@physics.mcgill.ca}
\affiliation{Department of Physics, McGill University, Montr\'eal, Qu\'ebec, H3A
2T8, Canada}

\author{Yi-Fu Cai}
\email{yifucai@physics.mcgill.ca}
\affiliation{Department of Physics, McGill University, Montr\'eal, Qu\'ebec, H3A
2T8, Canada}

\author{Robert H. Brandenberger}
\email{rhb@hep.physics.mcgill.ca}
\affiliation{Department of Physics, McGill University, Montr\'eal, Qu\'ebec, H3A
2T8, Canada}


\begin{abstract}
We examine reheating in the two-field matter bounce cosmology.
In this model, the Universe evolves from a matter-dominated phase of contraction
to an Ekpyrotic phase of contraction before the nonsingular bounce.
The Ekpyrotic phase frees the model from unwanted anisotropies,
but leaves the Universe cold and empty of particles after the bounce.
For this reason, we explore two particle production mechanisms which take
place during the course of the cosmological evolution:
Parker particle production where the matter field couples only to gravity
and particle creation via interactions between the matter field and the bounce field.
Although we show that both mechanisms can produce particles in this model,
we find that Parker particle production is sufficient to reheat the Universe to high temperatures.
Thus there is {\it{a priori}} no need to add an interaction term to the Lagrangian of the model.
Still, particle creation via interactions can contribute to the formation of matter and radiation,
but only if the coupling between the fields is tuned to be large.
\end{abstract}

\maketitle

\section{Introduction}

Reheating (see e.g.\ \cite{ABCM} for a recent review) is an integral part of inflationary
cosmology. Without a mechanism to convert the energy residing in the inflaton field
at the end of inflation into regular matter, inflation would produce a universe devoid
of any regular matter and radiation, obviously a cosmological catastrophe. In
a similar way, an early universe cosmology alternative to inflation must contain
a mechanism which leads to regular matter and radiation at late times.

The matter production mechanisms will differ from model to model. In inflationary
cosmology, it is a weak coupling between the inflaton field and
the fields representing regular matter which leads to a parametric resonance
(``preheating") instability in the equation of motion of these matter fields
during the time after the end of inflation when the inflaton is oscillating about its
ground state\ \cite{TB, KLS1, STB, KLS2}. In ``string gas cosmology"\ \cite{BV}, it is the
annihilation of string winding modes into string loops which automatically
leads to the generation of matter and radiation at the end of the initial
string phase (the ``Hagedorn" phase \textemdash\ see e.g.\ \cite{SGrev} for recent reviews
on string gas cosmology). In the ``emergent Galileon cosmology"\ \cite{Creminelli}
it is a ``defrosting" transition analogous to the resonant instability at the end
of inflation which leads to the production of regular matter\ \cite{Laurence}. This
mechanism also requires a coupling between regular matter fields and
the Galileon condensate.

A nonsingular bouncing cosmology with a matter-dominated initial
phase of contraction (during which scales of cosmological interest exit
the Hubble radius) is another alternative to cosmological inflation for
producing the spectrum of cosmological fluctuations observed today
\cite{Wands, Fabio}. The nonsingularity of the bounce can be obtained
by introducing new physics in the matter sector such as a quintom field
\cite{quintombounce} or a ghost condensate\ \cite{Chunshan}, or by
modifying gravity at high scales, e.g. in the context of string theory
\cite{SGbounce}, Horava-Lifshitz
gravity\ \cite{HLbounce}, nonlocal gravity\ \cite{Biswas}, or loop quantum
cosmology\ \cite{LQCbounce}. If there is no additional phase of
contraction during which the relative contribution of radiation to
the total energy density decreases, then there is no need for a
matter generation mechanism during and after the bounce since
the original matter and radiation content of the Universe at early
times during the contracting phase is preserved. However, it
is precisely such bouncing models without a dilution mechanism
during contraction which suffer from the ``anisotropy problem"
[``Belinsky-Khalatnikov-Lifshitz (BKL) instability''\ \cite{BKL};
the energy density in the anisotropies grows faster than the
energy density in matter and radiation and destroys the homogeneous
bounce]. To solve this anisotropy problem, a model with a phase
of Ekpyrotic contraction was introduced\ \cite{Cai1, Cai2} and it
was shown explicitly to be stable against anisotropies in\ \cite{Peter}
(see\ \cite{Cai:2014bea} for a review).
However, during the period of Ekpyrotic contraction the regular matter
and radiation become irrelevant analogously to what happens to
preinflationary matter and radiation during inflationary expansion,
and hence a mechanism is required to recreate matter and radiation
during and after the bounce.
This is the topic we address in the present study.

In this paper, we study gravitational matter field particle creation
in the two-field Ekpyrotic matter bounce model of\ \cite{Cai2}. This
mechanism is called ``Parker particle production"\ \cite{Parker:1968mv,Parker:1969au,Parker:1971pt}.
Due to the nontrivial dynamics of the matter mode functions in the
evolving background space-time, we find
that particles of matter fields can be efficiently generated. We evaluate
the particle number at the moment of matter-Ekpyrotic equality, and at
the beginning and end of the bouncing phase.
Our results explicitly show that by the time of the end of the
bouncing phase there is sufficient gravitational particle
production, and that there is hence no need to introduce any preheating
phase. This is very different from the situation in inflationary cosmology
where gravitational particle production is negligible compared to the
energy transfer via the preheating instability.
We further show that even if we introduce a small coupling between the
scalar field leading to Ekpyrotic contraction and the regular matter
fields, then parametric resonance is in fact negligible.

The paper is organized as follows. In Sec. II, we briefly review the model of two-field
matter bounce cosmology proposed in Ref.\ \cite{Cai2}. In Sec. III, we
study gravitational particle creation of a regular scalar matter field $\psi$ by
tracking the background evolution. In particular,
we present the quantization process of the $\psi$ particles in the matter
contraction, Ekpyrotic contraction, and the bouncing phases,
respectively. Afterwards, we compute the Bogoliubov coefficients of the $\psi$
particles by following the background evolution, and we determine the magnitude
of the energy density in the produced particles. To double check the validity of the
analyses of Sec. III, we analyze in Sec. IV the same setup by computing the
backreaction of the $\psi$ fluctuations on the energy-momentum tensor.
Then in Sec. V, we turn on an interaction term coupling the matter field to
the field yielding the Ekpyrotic contraction, and we investigate particle production
through direct interactions. We conclude with a discussion in Sec. VI.
Throughout the paper we take the sign of
the metric to be $(+,-,-,-)$ and define the reduced Planck mass by $M_p = 1/\sqrt{8\pi G}$.

\section{The Two-Field Matter Bounce Model}\label{Sec:model}

The two-field matter bounce\ \cite{Cai2} is a simple toy model which yields
regular matter and can give both an Ekpyrotic contraction and a nonsingular
bounce. In this section, we briefly review the dynamics of this nonsingular bounce model.

Following\ \cite{Cai2}, we consider the Lagrangian of the model to be
\begin{eqnarray}
 {\cal L} \, = \, K(\phi, X) + G(X)\Box\phi + P(\psi, Y)~,
\end{eqnarray}
where $\phi$ is the scalar field responsible for both the Ekpyrotic phase of contraction and
the bounce, and $\psi$ is a second scalar field representing the regular matter component
responsible for the initial matter contraction phase.

For the scalar field $\phi$, we introduce a $K$-essence-type term $K$ which is a function of
$\phi$ and its kinetic term\ \cite{Kessence}
\be
 X \, \equiv \ \frac{1}{2}g^{\mu\nu}\partial_\mu\phi\partial_\nu\phi \, ,
\ee
and a Horndeski-type\ \cite{Horn} term $G\Box\phi$ with
\be
 \Box \, \equiv  \, g^{\mu\nu}\nabla_\mu\nabla_\nu
\ee
being the standard d'Alembertian operator. When the Universe is far
away from the bounce, the Lagrangian of $\phi$ takes the canonical form which is the
sum of a regular kinetic term $X$ and a potential $V(\phi)$. More specifically,
we choose
\be
 K(\phi, X) \, = \, M_p^2 [ 1 - g(\phi)] X + \beta X^2 - V(\phi) \,
\ee
and
\be
 G(\phi, X) \, = \, \gamma X \, ,
\ee
with $\beta$ and $\gamma$ being two positive-definite constants, and
\be
 g(\phi) \, = \,  \frac{2 g_0}{e^{- \sqrt{2 / p} \phi} + e^{b_g \sqrt{2 / p} \phi}} \,  ,
\ee
where $g_0 > 1$, $p$, and $b_g$ are further positive constants. If the potential is chosen
to be a negatively valued exponential function, one can get a phase of Ekpyrotic contraction
\cite{Ekp} which will dilute the unwanted anisotropies\ \cite{noBKL}. The choice made
in\ \cite{Cai2} for the potential was
\be
 V(\phi) \, = \,  \frac{2 V_0}{e^{- \sqrt{2 / q} \phi} + e^{b_V \sqrt{2 / q} \phi}}  \,  ,
\ee
where $V_0$, $q$, and $b_V$ are more positive constants.
As a result, this type of model can avoid the BKL instability. The initial conditions
for $\phi$ will be taken to be a large negative value with small time derivative.
In this case, $\phi$ will contribute negligibly to the energy and momentum of
the background. As $\phi$ increases, it will eventually reach a value for which
$g(\phi) > 1$ and hence a ghost condensate\ \cite{ghostcond} will form for a short while.
This gives rise to a nonsingular bounce via the violation of the null energy condition.

Now we consider the dynamics of the massive scalar field $\psi$ which represents the
regular matter component. We take the Lagrangian to be of canonical form as
\begin{equation}
 P(\psi, Y) \, = \, Y-\frac{1}{2}m^2\psi^2~,
 \label{Lagrangianpsi}
\end{equation}
with $Y$ being defined as the kinetic term
\be
 Y \, \equiv \, \frac{1}{2}g^{\mu\nu}\partial_\mu\psi\partial_\nu\psi
\ee
and $m$ being the mass of the scalar. When $\psi$ oscillates around its vacuum
$\langle\psi\rangle=0$, its contribution to the energy-momentum tensor
is that of a pressureless fluid. If it dominates the total energy-momentum tensor,
then it leads to a matter-dominated evolution.

\begin{figure}
\includegraphics[scale=0.6]{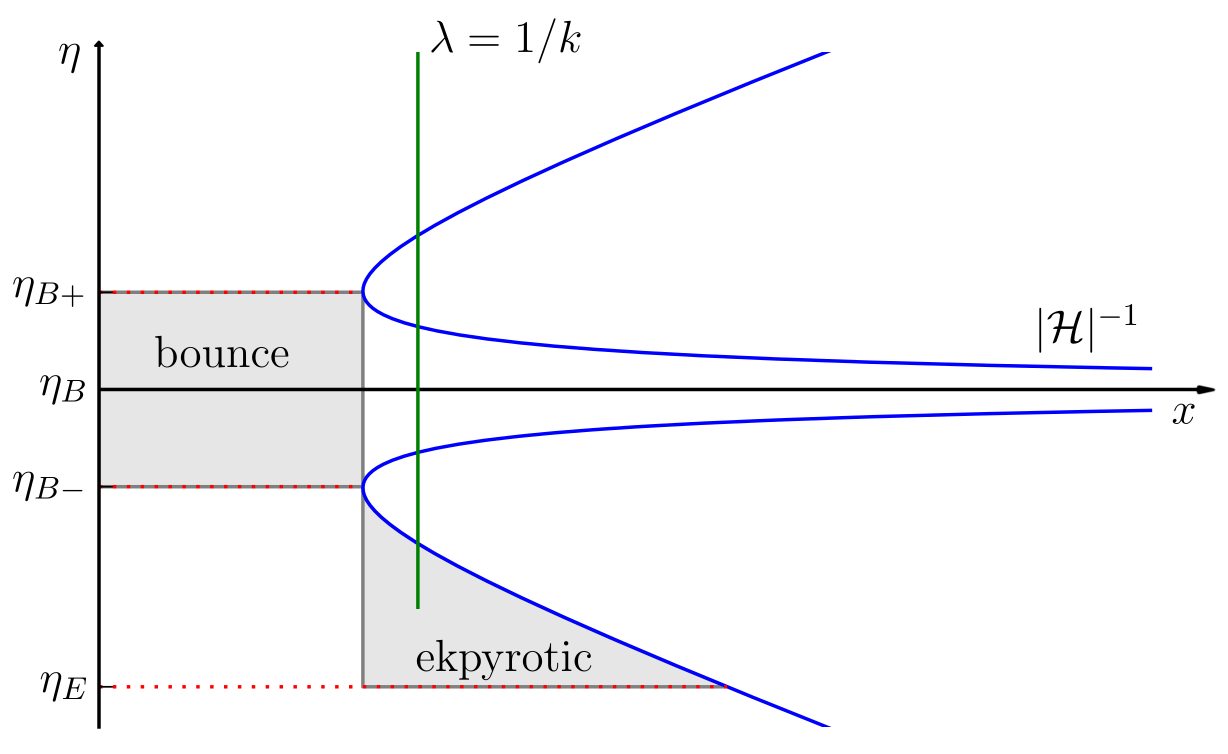}
\caption{Space-time sketch in the two-field matter bounce scenario. The comoving length and conformal time are labeled by $x$ and $\eta$, respectively. The solid blue curve shows $\mathcal{H}^{-1}$, the comoving Hubble radius.
The different phase transition times are labeled on the vertical axis. The shaded areas show the regions of integration for the respective phases (the Ekpyrotic and bounce phases are shown).
The comoving wavelengths of the fluctuating modes that we integrate originate in these shaded areas.
The green line shows the comoving wavelength of such a fluctuating mode that forms in the Ekpyrotic phase of contraction. We see that this mode reenters the Hubble horizon at later times, after $\eta_{B+}$.}
\label{fig:intmodes}
\end{figure}

Let us take a universe filled with the above two fields and consider a contracting universe
with initial conditions in which the contribution of the Ekpyrotic field
to the energy-momentum tensor is negligible since $\phi$ is initialized to have large
negative value and small velocity. Hence, the Universe will undergo a
phase of matter-dominated contraction with an effective background equation of state
$w=0$. Since the energy density in the Ekpyrotic field increases faster than that in
$\psi$, the Universe will at some point enter a period of
Ekpyrotic contraction driven by the scalar field $\phi$, with $w$ evolving to $w\gg1$.
Afterwards, the Universe experiences
a nonsingular bounce due to the emergence of the ghost condensate phase. The ghost
condensate phase ends naturally once $\phi$ exceeds a critical positive value. At that
point traditional cosmological thermal expansion takes over. The space-time
diagram of our nonsingular bounce cosmology is sketched in Fig.\ \ref{fig:intmodes}, in which
the vertical axis denotes time, and the horizontal axis denotes a comoving space coordinate.
The bounce point is taken to be $t = 0$. The wavelength of one mode is
depicted (the vertical line), and the Hubble radius is shown. It decreases
in comoving coordinates during the contraction phase and then increases again
in the expanding phase.

Since in the two-field matter bounce model we use the Ekpyrotic phase of contraction
to dilute the undesired anisotropies, this phase also washes out the primordial matter particles.
Therefore, one must be concerned that the Universe might be empty of regular matter
and radiation after the bounce. As discussed in the Introduction, a similar issue arises
within the framework of inflationary cosmology. In the case of inflation, in
order to have efficient production of regular matter after the phase of exponential
expansion, one usually introduces an explicit coupling term between the field
$\phi$ which yields inflation and regular matter fields. In this case, already
at the level of quantum field perturbation theory one expects the production
of matter particles\ \cite{DL, AFW} after inflation has ended, i.e.\ in a separate phase
called the ``reheating period." However, it was observed in\ \cite{TB} (see also
\cite{DK}) that the initial transfer of energy to matter fields is dominated by a
parametric resonance instability. This ``preheating period" was further
analyzed in\ \cite{KLS1, STB, KLS2}.

The question now arises as to whether it is necessary to introduce an analogous
reheating phase in a nonsingular matter bounce scenario in order to explain the
origin of regular matter and radiation. It is obviously possible to introduce such
a mechanism, and this was done within the framework of nonsingular bouncing cosmologies
in order to enhance the amplitude of the scalar cosmological perturbations relative
to the tensor modes\ \cite{bouncepreheating}. However, since the scalar modes in fact
experience a substantial growth in amplitude during the bouncing phase (greater than
that of the tensor modes)\ \cite{Cai2} (see also\ \cite{bouncepert} for earlier work
on fluctuations in a nonsingular bouncing cosmology), it is not necessary to
invoke bounce preheating to boost the scalar fluctuations, and it would be
more elegant not to do so since the bounce preheating scenario requires additional
couplings (like inflationary reheating does).

In this present work, we analyze matter production in the two-field matter bounce model
and show that there is no need to introduce preheating. Gravitational particle production
is in fact sufficiently effective. Specifically, we consider the production of $\psi$ particles
in our cosmological background. We are interested in particles with momenta which
are large (compared to the Hubble rate in the postbounce universe) rather than the large
wavelength fluctuations which contribute to cosmological perturbations.

\section{Gravitational particle production throughout the cosmic evolution}

In an expanding universe the concept of the vacuum state of a quantum field
is not well defined. In Minkowski space-time quantum field theory we can expand
the field in creation and annihilation operators which are the operator coefficients
of the positive and negative frequency modes, respectively. The vacuum state
is then the state which is annihilated by all of the annihilation operators. In
curved space-time there is no global definition of positive and negative
frequency modes. A mode which to an initial observer looks pure positive
frequency will look like a combination of positive and negative frequency
modes for a later time observer. The late time observer thus sees the
state as containing particles. This is the idea of Parker particle production
(see\ \cite{Parker:1968mv, Parker:1969au, Parker:1971pt}).

In the following, we will study Parker production of regular
matter (i.e.\ $\psi$) particles in our nonsingular bouncing cosmology.
Particle production is particularly efficient for modes which have
a wavelength larger than the Hubble radius. For such modes, the
canonical field variable is frozen out and squeezed. If the squeezed
mode function which was initially positive frequency is expanded at late
times into plane wave modes, then it obtains a large effective negative
frequency component which corresponds to the production of
potential $\chi$ particles (the use of the word ``potential'' indicates
that we can only speak of true particles once the modes have
reentered the Hubble radius and started to oscillate again).
The mixing between the initial positive frequency modes and
the late time modes is given by Bogoliubov coefficients, and
their values give us the number of produced particles (see
\cite{BD} for a textbook treatment of quantum field theory in
curved space-time).

We will focus on modes which have wavelengths which already
at early times (the time when we want to compute the density
of regular matter particles) allow an interpretation of the field
excitations as particles (i.e.\ short wavelengths).
We set up these modes in their initial vacuum state and
compute the Bogoliubov coefficients at later times in order
to determine the number of produced particles. Integrating
over Fourier modes then lets us calculate the energy density
in the produced matter particles.

To begin, let us recall the basic equations we will use.
Consider a homogeneous, isotropic and spatially flat FLRW universe. The corresponding
background dynamics is governed by the energy density and the pressure of the different
scalar fields (with energy densities are pressures denoted by $\rho$ and $P$,
respectively) through the Friedmann equations, which are given by
\begin{align}
 H^2 &= \frac{1}{3M_p^2} \left( \rho_\phi + \rho_\psi \right) ~, \nonumber\\
 \dot{H} &= -\frac{1}{2M_p^2} \left( \rho_\phi+P_\phi + \rho_\psi+P_\psi\right) ~,
\end{align}
where $H\equiv \dot{a}/a$ is the Hubble rate of the Universe and the dot denotes
the derivative with respect to the cosmic time. In addition, the scalar fields satisfy
the continuity equation so that covariant energy conservation is guaranteed.
The continuity equations can be derived by varying the Lagrangian with respect to the scalar
fields. Namely, if we focus on the matter field $\psi$, the resulting equation of motion is
\be
 (\Box + m^2)\psi \, = \, 0 \, ,
\ee
which is the Klein-Gordon equation.

At the background level, the matter field $\psi$ is only a function of cosmic time. However,
in order to study the quantum dynamics of its particle generation, we include the gradient
term and hence write down the equation of motion of the field as follows,
\begin{eqnarray}
 \ddot\psi +3 H(t) \dot\psi - \frac{\nabla^2}{a^2}\psi+ m^2\psi \, = \, 0~,
\end{eqnarray}
with $\nabla^2 \equiv g^{ij}\nabla_i\nabla_j$. Note that, in the current model we have
ignored possible interactions between $\psi$ and other fields.
For now, we focus on particle production of the field $\psi$ only under gravitational effects.
This assumption is in agreement
with our starting point that all fields are weakly coupled.

For convenience, let us work in conformal time, which is related to the cosmic time via
$d\eta = a^{-1}(t)dt$. As is well known, the matter equation simplifies if written
in terms of the rescaled field $\chi$ defined via
\be
 \chi(\eta,\mathbf{x}) \, = \, a(\eta)\psi(\eta,\mathbf{x}) \, .
\ee
The equation of motion for $\psi$ is then easy to translate into an equation of motion for $\chi$.
Since the equation of motion is linear, each Fourier mode evolves independently.
It is thus convenient to consider the Fourier transformation on the field $\chi$,
\begin{eqnarray}
 \chi(\eta,\mathbf{x}) \, = \, \int\frac{d^3k}{(2\pi)^{3/2}}\,\chi_k(\eta)e^{i\mathbf{k}\cdot\mathbf{x}}~,
\end{eqnarray}
and then track the dynamics of each Fourier mode along with the evolution of the background universe.
The equation of motion for a Fourier mode of $\chi$ becomes
\begin{eqnarray}\label{eom_chi}
 \chi_k''+\omega_k^2(\eta) \chi_k \, = \, 0 \, ,
\end{eqnarray}
with
\begin{eqnarray}\label{omega_square}
\omega_k^2 \, \equiv \, k^2 +a^2m^2 -\frac{a''}{a} \, ,
\end{eqnarray}
where primes denote derivatives with respect to conformal time.
It is the final term in (\ref{omega_square}) which is responsible
for the squeezing of the modes.

\subsection{Dynamics of the variable $\chi_k$}

We will solve the equation of motion\ \eqref{eom_chi} phase by phase throughout the
cosmological evolution. Specifically, we divide the time line into
four separate phases: the initial matter-dominated phase of contraction,
the Ekpyrotic phase of contraction, the nonsingular bouncing phase, and finally the
fast roll phase of expansion. In each of the phases, different approximations
are applicable.

\subsubsection{Matter contraction}

We first study the dynamics of $\chi_k$ in the matter-dominated phase of contraction.
In this stage, the background scale factor evolves as
\begin{eqnarray}\label{sol_a_m}
 a(\eta) \simeq c_E (\eta-\tilde\eta_E)^2 ~,~{\rm with}~~ c_E \equiv \frac{a_E}{(\eta_E-\tilde\eta_E)^2}~,
\end{eqnarray}
where we introduce $a_E \equiv a(\eta_E)$ as the value of the scale factor at the end
of this phase (and correspondingly, at the beginning of the Ekpyrotic phase). The coefficient
$\tilde\eta_E\equiv \eta_E - 2/{\cal H}_E$ would correspond to the conformal time
at the Big Crunch singularity if the matter phase of contraction lasted forever. Its inclusion is
very useful in regularizing the detailed computation. Additionally, we have introduced the
conformal Hubble rate ${\cal H}\equiv a'/a$.

Applying the relation\ \eqref{sol_a_m} to the definition of the effective frequency
square\ \eqref{omega_square}, one can derive
\begin{eqnarray}\label{omega_k_m}
 {\omega_k^m}(\eta)^2 = k^2 +c_E^2 m^2 (\eta-\tilde\eta_E)^4 -\frac{2}{(\eta-\tilde\eta_E)^2}~,
\end{eqnarray}
where the superscript ``$m$'' denotes the phase of matter contraction.
Then, by solving the key equation of motion\ \eqref{eom_chi},
one can determine the solution of $\chi_k$. At early times, i.e.\ for $\eta\ll\eta_E$,
it is obvious that the first two terms in
Eq.\ \eqref{omega_k_m} for the effective frequency are dominant.
Accordingly, the solution for the variable $\chi_k$ is approximately the plane wave
solution. Requiring vacuum initial conditions, we find the solution to be given by
\begin{eqnarray}\label{chi_m}
 \chi_k^m(\eta) \simeq \frac{e^{-i\omega_k^m (\eta-\tilde{\eta}_E)}}{\sqrt{2\omega_k^m}}~.
\end{eqnarray}

\subsubsection{Ekpyrotic contraction}

In the two-field matter bounce model, the potential of the scalar field $\phi$ is negative
and of exponential form. This kind of potential can drive $\phi$ into the attractor trajectory
for which the effective equation of state is larger than unity\ \cite{Ekp}.
As the Universe contracts, the energy density of the scalar field $\phi$ catches up with that of the
matter field $\psi$, and subsequently, a period of Ekpyrotic contraction begins. During this period,
the background scale factor can be expressed as a function of the conformal time in the
following approximate form
\begin{align}\label{sol_a_E}
 a(\eta) \simeq c_{B-} (\eta-\tilde{\eta}_{B-})^{\frac{q}{1-q}} ~,~~{\rm with}~
 c_{B-} \equiv \frac{a_{B-}}{(\eta-\tilde{\eta}_{B-})^{\frac{q}{1-q}}} ~,
\end{align}
and where $a_{B-} \equiv a(\eta_{B-})$ is the scale factor at the end of Ekpyrotic phase of
contraction (and equivalently, at the beginning of the bounce phase). Similar to the treatment
of the phase of matter contraction, we have introduced
$\tilde{\eta}_{B-}\equiv \eta_{B-}-\frac{q}{(1-q){\cal H}_{B-}}$ in the above expression.
Additionally, the parameter $q$ is associated with the background dynamics and is required
to be much less than unity so that the Ekpyrotic phase is achieved.

Making use of\ \eqref{sol_a_E}, one can easily obtain the squeezing term in the equation of
motion for $\chi_k$:
\begin{eqnarray}
 \frac{a''}{a} = -\frac{q(1-2q)}{(1-q)^2} \frac{1}{(\eta-\tilde\eta_{B-})^2}~.
\end{eqnarray}
The mass term $a^2m^2$ is subdominant since the scale factor has decreased to a small
value and the energy scale of the Universe has become higher than the mass scale $m$
after matter contraction. Therefore, the dominant part of Eq.\ \eqref{omega_square} is given by
\begin{eqnarray}\label{omega_k_c}
 {\omega_k^c}(\eta)^2 = k^2 +\frac{q(1-2q)}{(1-q)^2} \frac{1}{(\eta-\tilde\eta_{B-})^2}~,
\end{eqnarray}
where the superscript ``$c$'' denotes the phase of Ekpyrotic contraction.

Substituting the expression\ \eqref{omega_k_c} into the equation of motion\ \eqref{eom_chi},
one can solve the differential equation and find the explicit general solution to be
\begin{align}
 \chi_k^c(\eta) =& \sqrt{\eta-\tilde\eta_{B-}}\left\{ C_1(k)J_{\nu_c}\left[k\left(\eta-\tilde\eta_{B-}\right)\right]\right. \nonumber\\
 & \left.+C_2(k)Y_{\nu_c}\left[k\left(\eta-\tilde\eta_{B-}\right)\right]\right\} ~,
\end{align}
where $J_{\nu_c}$ and $Y_{\nu_c}$ are the two linearly independent Bessel functions
of the first and second kind, respectively. They have the same index
\be
 \nu_c \, = \, \frac{1-3q}{2(1-q)} \, \sim \, \frac{1}{2} \, .
\ee
For small values of the argument, i.e.\ for $|k(\eta-\tilde\eta_{B-})|<1$, the scalings
of the solutions are
\be
 \chi_k \sim \ (\eta-\tilde\eta_{B-})^{q} \,\, {\rm{and}} \,\, \chi_k \sim \ (\eta-\tilde\eta_{B-})^{1 - q} \, ,
\ee
which translates to
\be
 \psi_k \sim \ {\rm{const}} \,\, {\rm{and}} \,\, \psi_k \sim \ (\eta-\tilde\eta_{B-})^{1 - 2q} \, .
\ee
We see that there is one (almost) constant mode, while the second mode decays.

Modes in the large $k$ limit with $|k(\eta-\tilde\eta_{B-})|\gg1$ are still inside the Hubble radius
and keep the oscillating behavior. In this case, the solution can be simplified into the following
asymptotic form
\begin{align}\label{chi_c}
 \chi_k^c(\eta) \simeq \frac{1}{\sqrt{2 k}} \left[ \bar{C}_1(k) e^{ik(\eta-\tilde\eta_{B-})} 
 + \bar{C}_2(k) e^{-ik(\eta-\tilde\eta_{B-})} \right] ~,
\end{align}
with
\begin{align}
 \bar{C}_1 = \frac{2i+q\pi}{2\sqrt{\pi}} (C_1+iC_2)~,~~ \bar{C}_2 = \frac{-2i+q\pi}{2\sqrt{\pi}} (C_1-iC_2)~,
\end{align}
up to leading order. One can see that the $\bar{C}_1$ and $\bar{C}_2$ terms represent the negative
and positive frequency modes, respectively. If $\bar{C}_1$ is vanishing, the solution to $\chi_k$ is
identified as the vacuum fluctuation. However, if $\bar{C}_1$ is not zero, there will be particle production
from the vacuum state.

\subsubsection{Bouncing phase}

The phase of Ekpyrotic contraction stops when the scalar field $\phi$ evolves into the ghost
condensate state. Afterwards, the Universe experiences a nonsingular bouncing phase.
To quantitatively characterize this phase, it is useful to approximately parametrize the Hubble
parameter as a linear function of the cosmic time
\begin{eqnarray}
 H(t) = \Upsilon t~,
\end{eqnarray}
where $\Upsilon$ is a positive coefficient that determines how fast the bounce takes place.
We denote the beginning and the end of the bounce phase by $t_{B-}$ and $t_{B+}$, respectively.

Using conformal time, we can solve for the scale factor near the bounce,
\begin{eqnarray}
 a(\eta) = a_B +\frac{1}{2}a_B^3\Upsilon\eta^2 + \mathcal{O}(\Upsilon^2\eta^4)~,
\end{eqnarray}
under the fast bounce assumption (i.e.\ $\Upsilon \eta^2 < 1$ during the bouncing phase).
We note that it is convenient to set the bounce point
at $t_B=\eta_B=0$. We can then write down, up to leading order, the effective frequency
square in the bounce phase,
\begin{eqnarray}\label{omega_k_b}
 \omega_k^b(\eta)^2 = k^2 + a_B^2 (m^2-\Upsilon)~,
\end{eqnarray}
where the superscript ``$b$'' denotes the bounce phase. In this case, the solution can
be simply expressed as follows,
\begin{align}\label{chi_b}
 \chi_k^b(\eta) \simeq \frac{1}{\sqrt{2 \omega_k^b}} \left[ D_1(k) e^{ik \eta} + D_2(k) e^{-ik \eta} \right] ~,
\end{align}
again for the large $k$ modes.

\subsubsection{Fast roll expansion}

After the bounce, in the absence of backreaction of the fluctuations produced before and during the bounce,
the Universe would experience a period of fast roll expansion with an
effective equation of state $w=1$. The scale factor then evolves as
\begin{align}\label{sol_a_FR}
 a(t) = c_{B+} ( \eta-\tilde{\eta}_{B+} )^{\frac{1}{2}} ~,~~{\rm with}~
 c_{B+} \equiv \frac{a_{B+}}{( \eta_{B+}-\tilde{\eta}_{B+} )^{\frac{1}{2}}}~.
\end{align}
Analogously as before, $a_{B+}$ is the value of the scale factor after the bounce phase (and,
correspondingly, at the beginning of this phase of fast roll expansion). Also, we define
$\tilde{\eta}_{B+} \equiv \eta_{B+} - \frac{1}{2{\cal H}_{B+}}$, which
would correspond to the conformal time at the big bang if there were no bounce.
The above equation yields
\begin{eqnarray}\label{sol_app}
 \frac{a''}{a} = -\frac{1}{4(\eta-\tilde{\eta}_{B+})^2} ~,
\end{eqnarray}
in the phase of fast roll expansion.

One can substitute the scale factor\ \eqref{sol_a_FR} and the derived relation\ \eqref{sol_app}
into the expression\ \eqref{omega_square}
to determine the effective frequency square in the stage of fast roll expansion. We find that
the mass term is negligible
when the Hubble parameter of the Universe $H$ is larger than the mass of the matter field
$m$. As a result, we obtain the dominant part
of the effective frequency squared:
\begin{eqnarray}\label{omega_k_e}
 \omega_k^e(\eta)^2 = k^2 + \frac{1}{4(\eta-\tilde{\eta}_{B+})^2} ~.
\end{eqnarray}
As usual, the superscript ``$e$'' denotes the phase of fast roll expansion.

Making use of\ \eqref{omega_k_e}, the equation of motion\ \eqref{eom_chi} yields the following
solution,
\begin{align}
 \chi_k^e(\eta) =& \sqrt{\eta-\tilde\eta_{B+}} \left\{ E_1(k)J_{0}[k(\eta-\tilde\eta_{B-})]\right. \nonumber\\
 & \left.+E_2(k)Y_{0}[k(\eta-\tilde\eta_{B-})] \right\} ~,
\end{align}
where $E_1$ and $E_2$ are two coefficients to be determined later. Again, we focus on
the regime of large $k$ modes ($|k(\eta-\tilde\eta_{B+})|\gg1$) and can get the simplified form,
\begin{align}\label{chi_e}
 \chi_k^e(\eta) \simeq \frac{1}{\sqrt{2 k}} \left[ \bar{E}_1(k) e^{ik(\eta-\tilde\eta_{B+})}
 + \bar{E}_2(k) e^{-ik(\eta-\tilde\eta_{B+})} \right] ~,
\end{align}
with
\begin{align}
 \bar{E}_1 &= \frac{2i+\pi/3}{2\sqrt{\pi}} (E_1+iE_2)~, \nonumber \\
 \bar{E}_2 &= \frac{-2i+\pi/3}{2\sqrt{\pi}} (E_1-iE_2)~,
\end{align}
up to leading order.

\subsection{Particle production}\label{sec:particleprod}

We are interested in whether particles of the matter field have been generated
in the process of the evolution during the above different phases. As is
conventionally done in studying quantum field theory in curved space-time,
we work in the Heisenberg representation in which the operators carry
the time dependence.

To study particle production, it is important to express the solution of the field variable
$\chi_k$ at a given time $\eta$ as a sum of local positive and negative frequency
modes as follows,
\begin{align}
 \chi_k = \frac{1}{\sqrt{2\omega_k}} [ \alpha_k e^{-i\int^\eta \omega_k d\tilde\eta} + \beta_k e^{i\int^\eta \omega_k d\tilde\eta} ] ~,
 \label{sumposnegmodes}
\end{align}
where the coefficients $\alpha_k$ and $\beta_k$ are the so-called Bogoliubov coefficients.
If the mode functions evolve
adiabatically, then the Bogoliubov coefficients are time independent and
no particles are produced.

Following the standard approach of canonical quantization, one can define
the conjugate momentum $\Pi_k$ for the field variable through
\begin{eqnarray}
 \Pi_k \equiv \frac{\delta {\cal S}}{\delta\chi_k} = \chi_k'~.
\end{eqnarray}

Note that the general solutions to the field variable in different phases have been obtained in the
expressions\ \eqref{chi_m},\ \eqref{chi_c},\ \eqref{chi_b}, and\ \eqref{chi_e}, respectively. In
each phase, there are two modes. Given initial conditions at the beginning of the dynamics,
the actual solution in each phase is the combination of the two fundamental solutions in
that phase which is obtained by demanding the solutions to the variable $\chi$ and its
conjugate momentum (and hence its time derivative)
to be continuous across the transition between the previous phase and the one under
consideration. Given the solution in any given phase, we can then determine the Bogoliubov
coefficients. In doing this, we interpret the ``amount" of negative frequency modes
that are created during the entire evolution as the particle production.

We wish to consider the evolution of the vacuum state of $\chi_k^{\mathrm{vac}}(\eta)$
defined at some initial time $\eta_i$. First of all, we expand the field into combinations
$\chi_{k}^{{\rm{vac}}}$ and $\chi_{k}^{{\rm{vac}} *}$ of the two
fundamental solutions of the  classical mode equation which are positive and negative
frequency at $\eta_i$, respectively. After quantization, the coefficients $a_{\mathbf{k}}$ and
$a_{\mathbf{k}}^{\dag}$ become operators obeying the canonical commutation relations,
which in turn implies that the  $\chi$ field and its conjugate momentum $\Pi$ are canonically
conjugate operators. It is then possible to write the mode expansion
of the $\chi$ field as
\begin{equation}
 \chi(x)=\int\frac{d^3k}{(2\pi)^{3/2}}\left(a_{\mathbf{k}}\chi_k^{\mathrm{vac}}e^{i\mathbf{k}\cdot\mathbf{x}}
 +a_{\mathbf{k}}^{\dag}\chi_k^{\mathrm{vac}*}e^{-i\mathbf{k}\cdot\mathbf{x}}\right)~,
\end{equation}
where $a_{\mathbf{k}}^{\dag}$ and $a_{\mathbf{k}}$ represent the creation and annihilation
operators, respectively. The time dependence of the field operator is manifested by the time
dependence of the classical mode functions.

At a later time $\eta_{\rm{new}}$ we must expand the field operator $\chi$ in a new basis
of mode functions which are locally positive and negative frequency at the new time. Let us
denote the positive frequency mode at time $\eta_{\rm{new}}$ by
$\chi_k^{\mathrm{new}}(\eta)$. The field $\chi$ can also be expanded in terms of the new
basis modes
\begin{equation}
 \chi(x)=\int\frac{d^3k}{(2\pi)^{3/2}}\left(b_{\mathbf{k}}\chi_k^{\mathrm{new}}e^{i\mathbf{k}\cdot\mathbf{x}}
 +b_{\mathbf{k}}^{\dag}\chi_k^{\mathrm{new}*}e^{-i\mathbf{k}\cdot\mathbf{x}}\right)~,
\end{equation}
where $b_{\mathbf{k}}^{\dag}$ and $b_{\mathbf{k}}$ are new creation and annihilation operators.
Since $\chi_k^{\mathrm{vac}}$ and $\chi_k^{\mathrm{new}}$ form complete sets of mode solutions
at early and late times, it is possible to relate one set of modes to the other via
\begin{equation}
 \chi_k^{\mathrm{new}}=\alpha_k\chi_k^{\mathrm{vac}}+\beta_k\chi_k^{\mathrm{vac}*} \, ,
\end{equation}
where the coefficients are the famous Bogoliubov coefficients.
From this, it is straightforward to derive that the expectation value of the vacuum number
operator evaluated after the phase transition is equal to
\begin{equation}
 n_k=|\beta_k|^2~.
\end{equation}
This is the result that we will exploit throughout this work and interpret as the particle
number in mode $k$ which has been produced between the initial time and the time
$\eta_{\rm{new}}$.

\subsubsection{Particle number at the beginning of the Ekpyrotic phase}

We start our modes in their vacuum state during the matter phase of
contraction. At the time $t_{E}$, the equation of state transits into an
Ekpyrotic one. As we found in the previous subsections,
the fluctuating mode solutions $\chi_k$ in the matter contracting phase and
Ekpyrotic contraction period are given by
\eqref{chi_m} and\ \eqref{chi_c}, respectively. To relate the two in the limit of
small $q$, we can apply the matching condition by requiring
the solutions and their conjugate momenta to be continuous at the
transition point. This then yields up to leading order the following relation:
\begin{align}
 \beta_k \simeq& \frac{e^{-i\omega_k^c(\eta_E)\eta_E-i\omega_k^m(\eta_E)(\eta_E-\tilde\eta_E)}}{8\sqrt{3}k^2} \\ \nonumber
 &\times\left(\frac{2q(\tilde\eta_{B-}+\eta_E)}{(\tilde\eta_{B-}-\eta_E)^3}-10m^2c_E^2(\eta_E-\tilde\eta_E)^4\right)~,
\end{align}
for fluctuation modes with comoving wave number $k$ larger than the comoving Hubble parameter ${\cal H}_E$.
The number of particles that is then produced is
\begin{align}\label{nktE}
 n_k=|\beta_k|^2\simeq&\frac{1}{192k^4}\Big[100m^4c_E^4(\eta_E-\tilde\eta_E)^8 \\ \nonumber
 &-\frac{40qm^2c_E^2(\eta_E-\tilde\eta_E)^4(\tilde\eta_{B-}+\eta_E)}{(\tilde\eta_{B-}-\eta_E)^3}\Big]~.
\end{align}

To evaluate the contribution of these new particles to the background evolution, one
should compute the energy density of these particles
and compare it to the background energy density. Let us take for granted the fact that
the energy density of the particles that have been produced is given by
\begin{equation}
 \rho_{\chi}=\frac{1}{(2\pi)^3a^4}\int d^3k\,n_k\omega_k\simeq\frac{1}{2\pi^2a^4}\int dk\,k^3n_k~,
 \label{rhochi}
\end{equation}
in the limit where $\omega_k\simeq k$.
We will justify that this is the form that the energy density of particle creation should
take in Sec.\ \ref{sec:back-reaction}. However, similar to the vacuum energy density,
it is easy to realize that this integral diverges, both in the ultraviolet (UV) and in the
infrared (IR). For this reason, it is necessary to introduce some cutoffs. First, all of
the above analysis has been done for modes on sub-Hubble scales at the initial time. Thus, the
comoving scale corresponding to the initial Hubble radius yields the natural IR cutoff.
Second, we impose an UV cutoff that  corresponds to the Hubble scale at the transition
between matter contraction and Ekpyrotic contraction. In other words, we integrate modes
with value of $k$ smaller than $\mathcal{H}_E$. When considering particle production
at the beginning of the bounce phase, we need to consider modes with $k$
between $\mathcal{H}_E$ and $\mathcal{H}_{B-}$. During the bounce phase, we take the
Planck scale to be the UV cutoff, so we integrate modes with scale
between $\mathcal{H}_{B-}$ and $a_B M_p$ (see Fig.\ \ref{fig:intmodes} for a pictorial representation).

Using Eq.\ \eqref{nktE} for the number of particles created at $t_E$, performing the
integral\ \eqref{rhochi}, and dividing the resulting energy density by the background
energy density $\rho_{\mathrm{back}}\simeq 3M_p^2H_E^2$,
we find the following result for the energy density of the particles produced
during the matter phase of contraction
(as a fraction of the background energy density):
\begin{align}\label{rhochi_m-e}
 \frac{\rho_{\chi}}{\rho_{\mathrm{back}}}\simeq&\frac{m^2}{336\pi^2M_p^2H_E^2}\left[30m^2-7q\frac{(\tilde\eta_{B-}+\eta_E)}{(\tilde\eta_{B-}-\eta_E)^3}\right] \\ \nonumber
 &\times\ln\left(\frac{a_{B-}H_{B-}}{k_{\mathrm{min}}}\right)~,
\end{align}
where $k_{\mathrm{min}}$ represents the comoving wave number of the fluctuating
mode with the largest wavelength.

Let us explore the consequences of Eq.\ \eqref{rhochi_m-e}. Overall, the energy density
of particle production after the matter phase of contraction is very small compared
to the background energy density. The prefactor of Eq.\ \eqref{rhochi_m-e} can be of
order 1, but the square bracket is suppressed due to the mass squared term and the
second term is suppressed due to the large
duration of the Ekpyrotic phase. The latter is necessary to dilute unwanted anisotropies
(see Ref.\ \cite{Cai2}). Finally, the log term could be arbitrarily large for arbitrarily
small values of $k_{\mathrm{min}}$, but recalling that the fluctuations
must reenter the Hubble radius after the bounce, this term cannot contribute to the
energy density significantly.

\subsubsection{Particle number at the beginning of the bounce phase}

Using the same technique as above, we require the mode solutions $\chi_k$ given by
Eqs.\ \eqref{chi_c} and\ \eqref{chi_b} to be continuous at the transition time $t_{B-}$.
The matching conditions then yield the following Bogoliubov coefficient:
\begin{equation}
 \beta_k\simeq e^{-ik(\eta_{B-}-\tilde\eta_E)-i\omega_k^b\eta_{B-}}\frac{\omega_k^b-k}{2\sqrt{3k\omega_k^b}}~,
\end{equation}
where we recall that the effective frequency in the bounce phase is given by Eq.\ \eqref{omega_k_b}.
Up to leading order in $k$, we find the resulting number of particles to be given by
\begin{equation}
 n_k=|\beta_k|^2\simeq\frac{a_B^4(m^2-\Upsilon)^2}{48k^4}~.
\end{equation}
Using the fact that $m^2\ll\Upsilon$ and, since the bounce phase is short, assuming
that $a(\eta_{B-})\simeq a_B$, the energy density of particle production is found to be
\begin{equation}
 \rho_{\chi}\simeq\frac{\Upsilon^2}{96\pi^2}\ln\left(\frac{a_{B-}H_{B-}}{a_EH_E}\right)~,
\end{equation}
and in terms of a fraction of the background energy density, we find
\begin{equation}\label{rhochi_e-b}
 \frac{\rho_{\chi}}{\rho_{\mathrm{back}}}\simeq\frac{1}{288\pi^2M_p^2t_{B-}^2}\ln\left(\frac{a_{B-}H_{B-}}{a_EH_E}\right)~.
\end{equation}
Note that we used the relation $H_{B-}=\Upsilon t_{B-}$.

This last result shows that gravitational particle production starts to become important
in the Ekpyrotic phase of contraction. If the bounce phase is very fast (a few Planck
times for instance), then the energy density of particle production is
of the order of $\mathcal{O}(10^{-4})$ times the background energy density. However,
recent studies suggest that the bounce phase should last about $10^3-10^4$ Planck
times to obtain the correct amplitude of cosmological perturbations and a
sufficiently small tensor-to-scalar ratio (\cite{Cai:2014xxa,Battarra:2014tga}), which
reduces the fraction (\ref{rhochi_e-b}).

\subsubsection{Particle number at the end of the bounce phase}

For the transition in the equation of state occurring at $t_{B+}$, we require continuity of
the mode solutions given by Eqs.\ \eqref{chi_b} and\ \eqref{chi_e}.
Using the same techniques as before, we find the Bogoliubov coefficient $\beta_k$ and
the resulting number of particles. The latter is found to be
\begin{equation}
 n_k\simeq\frac{a_B^4\Upsilon^2}{48k^4}(4\Upsilon^2t_{B+}^4-\Upsilon t_{B+}^2-1)^2~,
\end{equation}
where we used the facts that $m^2\ll\Upsilon$, $a(\eta_{B+})\simeq a_B$, and
$\eta_{B+}\simeq t_{B+}/a_{B+}$.
The resulting energy density in terms of the background energy density is
found to be
\begin{equation}\label{rhochi_b-f}
 \frac{\rho_{\chi}}{\rho_{\mathrm{back}}}\simeq\frac{(4\Upsilon^2t_{B+}^4-\Upsilon t_{B+}^2-1)^2}{288\pi^2M_p^2t_{B+}^2}\ln\left(\frac{a_BM_p}{a_{B-}H_{B-}}\right)~.
\end{equation}
Here again, we see that particles have been produced. In fact, we see that
Eqs.\ \eqref{rhochi_e-b} and\ \eqref{rhochi_b-f} share the same
denominator if one assumes that the bounce phase is symmetric ($|t_{B-}|=|t_{B+}|$).
However, the energy density after the bounce phase
has an extra factor which tends to 1 as the bounce duration tends to 0.
So, in the short bounce limit, the energy densities of particle
production at the end of the Ekpyrotic phase and at the end of the bounce phase
are comparable, whereas for a longer bounce, the contribution from
the bounce phase dominates. At this point, the energy density of particle production
is at most $\mathcal{O}(10^{-3})$ times the background energy density.

\subsection{Reheating time and reheating temperature}

We found in the previous section that particles are gravitationally produced throughout
the evolution of the Universe, but that the dominant contributions came from late in the
Ekpyrotic phase of contraction and from the bounce phase. We also found that the energy
density of these particles could reach about $10^{-3}$ times the background energy density
at the end of the bounce phase.
On one hand, this is not enough to disturb the background evolution,
i.e., it does not lead to large backreaction effects. On the other hand, this is enough to
reheat the Universe since after the bounce phase, particles that have been produced will
redshift less fast than the background field and will hence ultimately become dominant.

Specifically, we know that $\rho_{\chi}\sim a^{-4}$, whereas $\rho_{\mathrm{back}}\sim a^{-6}$.
Defining the ``reheating time" $t_R$ as the time when the total energy density of particle
production equals the background energy density,
\begin{equation}
 \rho_{\chi,\mathrm{total}}(t_R)=\rho_{\mathrm{back}}(t_R)~,
 \label{tReq}
\end{equation}
we find that reheating occurs at
\begin{equation}
 t_R=\tilde t_{B+}+\frac{M_p^3a_{B+}^6}{(3\rho_*)^{3/2}(t_{B+}-\tilde t_{B+})^{2}}~,
\end{equation}
where
\begin{align}
 \rho_*=&\frac{\Upsilon^2a_{B+}^4}{96\pi^2}\left[\ln\left(\frac{a_{B-}H_{B-}}{a_EH_E}\right)\right. \\ \nonumber
 &\left.+(4\Upsilon^2t_{B+}^4-\Upsilon t_{B+}^2-1)^2\ln\left(\frac{a_BM_p}{a_{B-}H_{B-}}\right)\right]
\end{align}
is the sum of the energy densities from the Ekpyrotic phase and the bounce phase such that
\be
 \rho_{\chi,\mathrm{total}}(t_{B+}) \, = \, \rho_*a^{-4}(t_{B+}) \, .
\ee
Finally, using the relation
\be
 \rho_{\chi,\mathrm{total}}(t_R) \, = \, \frac{\pi^2}{15}T_R^4 \, ,
\ee
we find that the reheating temperature is given by
\begin{align}
 T_R=&\left(\frac{15}{9}\right)^{1/4}\frac{\rho_*^{3/4}}{\pi^{1/2}M_pH_{B+}a_{B+}^3} \\ \nonumber
 =&\frac{10^{1/4}}{48\pi^2}\frac{\Upsilon^{3/2}}{M_pH_{B+}}\left[\ln\left(\frac{a_{B-}H_{B-}}{a_EH_E}\right)\right. \\ \nonumber
 &\left.+(4\Upsilon^2t_{B+}^4-\Upsilon t_{B+}^2-1)^2\ln\left(\frac{a_BM_p}{a_{B-}H_{B-}}\right)\right]^{3/4}~.
\end{align}
For a short bounce, this temperature can be of the order of $\mathcal{O}(10^{-5})M_p$,
about an order of magnitude below the GUT scale. Using parameter values that better suit the recent
observations (\cite{Cai:2014xxa}), we find $T_R\sim\mathcal{O}(10^{-7})M_p$.

Let us briefly compare reheating in inflationary cosmology and the bounce reheating mechanism
by Parker particle production discussed here. Neither Parker particle production nor inflationary
preheating produce a state with a thermal distribution of particles at the time that matter starts
to dominate the energy density. A period of thermalization (e.g.\ via perturbative processes) is
required both in inflation and in our bouncing cosmology. In fact, the processes involved will
be identical. If we compare the energy density at which regular matter starts to dominate,
then the value $T_R\sim\mathcal{O}(10^{-7})M_p$ in our scenario is comparable to the
``temperature" after preheating in intermediate energy inflation models. For a review
of the challenges of the actual thermalization process the reader is referred to\ \cite{ABCM}.

\section{Backreaction of Thermal Particles from Cosmic Fluctuations}\label{sec:back-reaction}

In this section we take another view on the backreaction of the produced particles on
the background dynamics. Instead of focusing directly on the number of produced
particles and computing their associated energy density we will consider the energy-momentum
tensor of the scalar field and directly compute its backreaction.

We know that for a scalar field $\psi$ with a canonical Lagrangian given by
Eq.\ \eqref{Lagrangianpsi}, the energy density  due to modes with $k\gg\mathcal{H}$
is given by
\begin{equation}
 \rho_{\mathrm{br}}\simeq\frac{1}{2}\langle(\delta\dot\psi)^2\rangle
 +\frac{1}{2a^2}\langle(\nabla\delta\psi)^2\rangle+\frac{1}{2}m^2\langle(\delta\psi)^2\rangle~,
 \label{rho1}
\end{equation}
where $\delta\psi$ denotes the fluctuation of the scalar field. Note that in contrast
to a field $\phi$ responsible for Ekpyrotic contraction, $\psi$ has no homogeneous
nonvanishing background value. In the above expectation values, only sub-Hubble
Fourier modes are considered. If we wanted to discuss the backreaction of
super-Hubble modes, it would be essential to include the metric fluctuations induced
by the matter modes. In the context of inflationary cosmology, the backreaction
formalism for long wavelength modes was developed in detail in Ref.\ \cite{Abramo:1997hu}.

Recalling that $\chi=a\psi$ and using the mode expansion of the $\chi$ field, it is
straightforward to express the above energy density as
\begin{align}
 \rho_{\mathrm{br}}\simeq&\frac{1}{(2\pi)^3a^4}\int d^3k\,\frac{1}{2}\Big(|\chi_k'|^2-\mathcal{H}\left(\chi_k'^*\chi_k+\chi_k'\chi_k^{*}\right) \nonumber \\
 &+\left(\mathcal{H}^2+k^2+m^2a^2\right)|\chi_k|^2\Big)\,.
\end{align}
Expanding the $\chi_k$ modes as sums of positive and negative frequency modes
[see Eq.\ \eqref{sumposnegmodes}] and using the fact that $\omega_k\simeq k$ in
the large $k$ limit, the backreaction energy density becomes
\begin{equation}
 \rho_{\mathrm{br}}\simeq\frac{1}{(2\pi)^3a^4}\int d^3k\,\frac{1}{2}\omega_k\left(|\alpha_k|^2+|\beta_k|^2\right)\,.
\end{equation}
Noting that $|\alpha_k|^2-|\beta_k|^2=1$ and $n_k=|\beta_k|^2$, we can
reexpress the last result as
\begin{equation}
 \rho_{\mathrm{br}}\simeq\frac{1}{(2\pi)^3a^4}\int d^3k\,\frac{1}{2}\omega_k+\frac{1}{(2\pi)^3a^4}\int d^3k\,\omega_kn_k\,.
\end{equation}
The first term of this equation is the (divergent) vacuum energy, which we ignore.
The second term is what we wished to show as being the energy density of particle
production and it is what we used throughout Sec.\ \ref{sec:particleprod}.

\section{Particle Production Through Direct Interactions}

In inflationary cosmology particle production through direct interactions between the
matter field and the inflaton field is much more important than gravitational particle
production. The energy density due to gravitational particle production is
of the order $H^4$, where $H$ is the Hubble expansion rate during inflation,
whereas the energy density transferred to matter via preheating is proportional
to $H^2 M_p^2$ which is parametrically larger by a factor of $(M_p / H)^2$.
Thus, an obvious question to ask is how large the contribution of direct
interactions to matter production is in our matter bounce scenario.

The efficiency of the parametric resonance instability\ \cite{TB} underlying
preheating can be traced\ \cite{KLS2} to the fact that the inflaton oscillates
many times through $\varphi = 0$. During each crossing, the adiabaticity
condition for the matter modes is violated, leading to many bursts
of particle production. In contrast, in our model $\phi$ crosses zero only
once, and so less particle production through direct interactions is
expected.

On the other hand, in the emergent Galileon cosmology of\ \cite{Creminelli},
particle production via direct interactions turns out to be very efficient in
spite of the fact that the dynamical background field crosses $\phi = 0$
only once\ \cite{Laurence}. However, as explained in\ \cite{Laurence},
this is unexpected and due to the particular couplings of the model.
Hence, taking the lessons of\ \cite{Laurence} into account we expect
a small amount of particle production via interactions. In the following,
we will verify this expectation.

To obtain direct particle production we need to add an interaction
Lagrangian to the problem (as one needs to in order to obtain inflationary
reheating). We take this coupling to be given by the last term in the
following Lagrangian
\begin{equation}
 \mathcal{L} \, = \, K(\phi,X)+G(\phi,X)\Box\phi+P(\psi,Y)-\frac{1}{2}\lambda^2\phi^2\psi^2M_p^2~,
\end{equation}
for some real coupling constant $\lambda$. We assume that the coupling
only takes place during the bounce phase where we expect particle production to be
most significant. Outside the bounce phase, we assume that the coupling constant $\lambda$ is very small.

Using the above Lagrangian with the interaction term, the equation of motion (EoM) for
$\psi$ in Fourier space becomes
\begin{equation}
 \ddot\psi_k+3H\dot\psi_k+\left(\frac{k^2}{a^2}+m^2+\lambda^2\phi^2M_p^2\right)\psi_k \, = \, 0~.
\end{equation}
Defining a new auxiliary field as $X_k(t)\equiv a^{3/2}(t)\psi_k(t)$, the equation of motion
simplifies and becomes
\begin{equation}
 \ddot X_k + \omega_k^2X_k \, = \, 0~,
 \label{eomX}
\end{equation}
where the time-dependent effective frequency is given by
\begin{equation}
 \omega_k^2 \, = \, \frac{k^2}{a^2}+m^2+\lambda^2\phi^2M_p^2-\frac{9}{4}H^2-\frac{3}{2}\dot H\,.
 \label{wk2}
\end{equation}
Since the background evolution in the bounce phase is well known (see\ \cite{Cai2}),
we could write down the form of
Eq.\ \eqref{wk2} explicitly, but we will be mainly interested in the effective frequency
close to the bounce point
where $\phi$ crosses 0. We note that
\begin{equation}
 \phi(t)\simeq\frac{\sqrt{\pi}}{2}\dot\phi_BT\mathrm{erf}\left(\frac{t}{T}\right)~,
\end{equation}
where $T$ is given by
\begin{equation}
 T\simeq\frac{H_{B+}}{\Upsilon}\sqrt{\frac{2}{\ln(\dot\phi_B^2/6H_{B+}^2)}}~.
\end{equation}
Thus, as long as the ratio $t/T$ remains small, it is sufficient to expend the
effective frequency squared to second order in time as follows,
\begin{align}
 \label{efffreq}
 \omega_k^2(t)\simeq&\left(\frac{k^2}{a_B^2}+m^2-\frac{3\Upsilon}{2}\right) \\ \nonumber
 &+\left(\lambda^2\dot\phi_B^2M_p^2-\frac{9\Upsilon^2}{4}-\frac{k^2\Upsilon}{a_B^2}\right)t^2 \\
 \nonumber \equiv& \, b + c t^2 \, ,
\end{align}
where we have denoted the zeroth-order term as $b$ and the coefficient
of the quadratic term as $c$.

To determine when particle production occurs, we need to find when the adiabaticity
condition is violated. This is the case when
\begin{equation}
 |\dot\omega_k|\gtrsim\omega_k^2~.
 \label{adiavio}
\end{equation}
Using Eq.\ \eqref{efffreq} for the effective frequency, the inequality\ \eqref{adiavio}
is equivalent to
\begin{equation}
 c^3t^6+3bc^2t^4+c(3b^2-c)t^2+b^3\lesssim 0~.
\end{equation}
Since the constant term in the above equation is negligible when $\lambda$ is large, we
can solve for $t$ and thus we
find that particle production occurs without interruption
for $t_i\lesssim t\lesssim t_f$ with
\begin{equation}
 t_f \, = \, \sqrt{\frac{\sqrt{4c-3b^2}-3b}{2c}}
\end{equation}
and $t_i=-t_f$. We note that this is only valid if $c>3b^2$, which translates into an upper
bound%
\footnote{As in Sec.\ \ref{sec:particleprod}, we assume $m^2\ll\Upsilon$.},
\begin{equation}
 k_{\mathrm{max}}=a_B\sqrt{\frac{4\Upsilon+\sqrt{3\lambda^2\dot\phi_B^2M_p^2-11\Upsilon^2}}{3}}~,
\end{equation}
as long as $\lambda>3\Upsilon/\dot\phi_BM_p$.

In order to solve the EoM, let us write down the solution in the form of
\bea
 X_k(t) \, &=& \, \frac{1}{\sqrt{2\omega_k(t)}} \\
 & & \times\left[ \alpha_k(t)e^{-i\int_{t_i}^td\tilde t\,\omega_k(\tilde t)}+\beta_k(t)e^{i\int_{t_i}^td\tilde t\,\omega_k(\tilde t)}\right]~. \nonumber
\eea
Its conjugate momentum is then given by
\bea
 \Pi_k(t) \, &=& \, i\sqrt{\frac{\omega_k(t)}{2}} \\
 & & \times\left[ -\alpha_k(t)e^{-i\int_{t_i}^td\tilde t\,\omega_k(\tilde t)}+\beta_k(t)e^{i\int_{t_i}^td\tilde t\,\omega_k(\tilde t)}\right]~, \nonumber
\eea
and the Bogoliubov coefficients satisfy the following equations:
\begin{eqnarray}
 \dot\alpha_k \, &=& \, \frac{\dot\omega_k}{2\omega_k}e^{2i\int_{t_i}^td\tilde t\,\omega_k(\tilde t)}\beta_k~, \\
 \dot\beta_k \, &=& \, \frac{\dot\omega_k}{2\omega_k}e^{-2i\int_{t_i}^td\tilde t\,\omega_k(\tilde t)}\alpha_k~.
\end{eqnarray}
To set the initial conditions, we require that there are no particles at $t_i$, which implies
that $\alpha_k(t_i)=1$ and $\beta_k(t_i)=0$.

Once we find the solution $X_k(t)$, we know that the total number of particles that
will have been produced is given by
\begin{equation}
 n_k(t_f) \, = \, \frac{\omega_k(t_f)}{2}\left(\frac{|\dot X_k(t_f)|^2}{\omega_k^2(t_f)}+|X_k(t_f)|^2\right)-\frac{1}{2}~,
\end{equation}
or equivalently by
\begin{equation}
 n_k(t_f) \, = \, |\beta_k(t_f)|^2~.
\end{equation}
An approximate solution is given by (see the Appendix for details)
\begin{equation}
 n_k(t_f) \, \simeq \, \sinh^2\left[\frac{t_f^2\sqrt{c}}{2}-\frac{b}{2\sqrt{c}}\ln\left(1+\frac{ct_f^2}{b}\right)\right]~.
\end{equation}
The energy density of particle production at the end of the bounce phase is then given by
\begin{equation}
 \rho_X(t_{B+})=\frac{1}{2\pi^2a(t_{B+})^4}\int_{0}^{k_{\mathrm{max}}} dk~k^2\omega_k(t_{B+})n_k(t_f)~,
 \label{rhoX1}
\end{equation}
where we used the fact that $n_k(t_{B+})=n_k(t_f)$ since particle production stops before
the end of the bounce phase.
As explained in the Appendix, the result for $n_k(t_f)$ is only valid
for $b>0$, which imposes a lower bound
$k_{\mathrm{min}}=a_B\sqrt{3\Upsilon/2}$. Since we are interested in the limit where
$\lambda\gg\Upsilon$,
we notice that $k_{\mathrm{max}}\gg k_{\mathrm{min}}$ meaning that the integral is
dominated by its UV cutoff.
Thus, we can effectively take the lower integration bound to be 0.

\begin{figure}[t]
\includegraphics[scale=0.68]{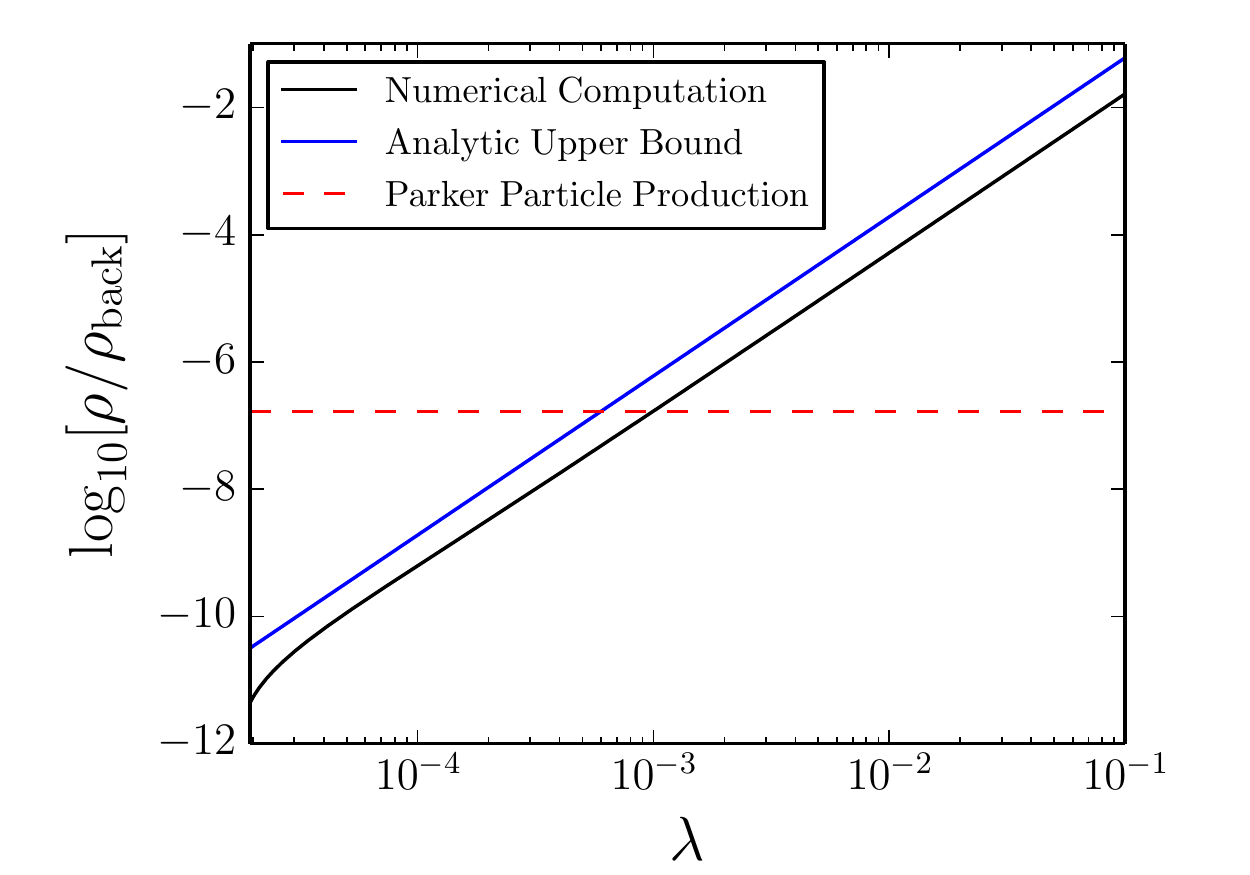}
\caption{Plot of the ratio of the energy density of particle production to the background energy density as
a function of the coupling constant. The slope of the Hubble parameter and the physical duration of the bounce are taken to be
$\Upsilon = 5.5 \times 10^{-7}\,M_p^2$ and $t_{B+} = 2.1 \times 10^3\,M_p^{-1}$, respectively. The blue curve shows the analytic upper bound
found in Eq.\ \eqref{rhoXf} and the black curve was obtained by solving the EoM and the integral for the energy density numerically.
The dashed red line is the contribution from Parker particle production that we found earlier [see Eq.\ \eqref{rhochi_b-f}].}
\label{fig:comp}
\end{figure}

The integrand of Eq.\ \eqref{rhoX1} is a complicated function of $k$, but it is
straightforward to find an upper bound to the overall
contribution. For large values of the coupling constant, we see from Eq.\ \eqref{wk2}
that the effective frequency squared
at the end of the bounce phase is dominated by the term $\lambda^2\phi(t_{B+})^2M_p^2$.
We also notice that $n_k(t_f)\lesssim\sinh^2(1/2)$ for all allowed values of $k$, so
Eq.\ \eqref{rhoX1} simplifies to become the inequality
\begin{equation}
 \rho_X(t_{B+}) \lesssim \frac{a_B^3(\lambda\dot\phi_BM_p)^{5/2}H_{B+}\mathrm{erf}(t_{B+}/T)\sinh^2(1/2)}{3^{7/4}(2\pi)^{3/2} a_{B+}^4 \Upsilon \sqrt{\ln(\dot\phi_B^2/6H_{B+}^2)}}~.
 \label{rhoXf}
\end{equation}
This can be compared to the energy density of Parker particle production
(recall Eq.\ \eqref{rhochi_b-f}),
\begin{align}
 \frac{\rho_X(t_{B+})}{\rho_{\chi}(t_{B+})} \lesssim &\frac{8\sqrt{2\pi} a_B^3 (\lambda\dot\phi_BM_p)^{5/2}t_{B+} \mathrm{erf} \left(\frac{t_{B+}}{T}\right) \sinh^2\frac{1}{2}}{3^{3/4}a_{B+}^4 \Upsilon^2\ln\left(\frac{a_BM_p}{a_{B-}H_{B-}}\right) \sqrt{\ln\left(\frac{\dot\phi_B^2}{6H_{B+}^2}\right)}} \\ \nonumber
 &\times (4\Upsilon^2t_{B+}^4-\Upsilon t_{B+}-1)^{-2}~.
\end{align}
From this result, we see that for a fixed energy scale $H_{B+}$ and a fixed
coupling constant $\lambda$, the ratio of the energy densities
tends to zero as $\Upsilon\rightarrow 0$ or as $\Upsilon\rightarrow\infty$.
This means that in the limit where the bounce is either infinitely long or infinitely
short, the energy density from Parker particle production will always dominate the
energy density from interaction.
However, for typical bounce durations, the ratio of the energy densities depends on
the size of the coupling constant.
An example is shown in Fig.\ \ref{fig:comp} where we see that particle production from
interaction can contribute significantly to Parker particle production, but only if
the coupling between the two fields is quite large.
In general, for smaller values of the coupling constant, the energy density from
Parker particle production dominates over
the energy density from interaction.

\section{Conclusions}\label{Sec:conclusions}

We have considered Parker particle production in the two-field
matter bounce model
of\ \cite{Cai1}, a prototypical example of a nonsingular matter bounce which can generate
a scale-invariant spectrum of cosmological perturbations and which is stable against
anisotropies near the bounce point. The stability to anisotropies is obtained via an
Ekpyrotic phase of contraction. This phase of Ekpyrosis, on the other hand, also washes
out any matter and radiation which might have existed early in the contracting phase.
Hence, to make the model viable, a reheating mechanism is required.

We have shown that in our background, gravitational Parker particle production is
sufficiently effective to reheat the Universe to high temperatures.
For minimal coupling between the scalar field $\phi$ yielding
the Ekpyrotic contraction and the nonsingular bounce and regular matter
the effect of particle production through interactions is generally small,
but could contribute to Parker particle production if the coupling were large.
We thus see that we do not need to introduce extra ingredients
into our matter bounce model in order to obtain a hot postbounce universe dominated
by regular matter and radiation. This contrast with the case of inflationary cosmology,
where the direct coupling of the inflaton field with matter
generically produces an energy density in the reheat
fields which is of the order of the inflationary
energy density, whereas the contribution of Parker
particle production is of the order $\rho \sim H^4$
and which is hence suppressed compared to the density
of particles produced by preheating by a factor of order
$(H / M_p)^2$. The reason for this difference is
related to the fact that in the case of inflation the
inflaton field loses most of its energy density during
the reheating process, whereas in our nonsingular
bounce the field $\phi$ retains most of its energy.

Finally, we note that the methods developed here are applicable to the study of Parker
particle production in other nonsingular bouncing models, e.g.\ in the new Ekpyrotic universe\ \cite{newEkp}.
Based on the arguments of the previous paragraph we conjecture that also in these other
models, Parker particle production will be sufficient to reheat the Universe,
and that Parker particle production will not be suppressed relative to particle production via
direct couplings. To support this conjecture, we point out that Parker particle production has been
shown to be efficient for certain models of the matter bounce in loop quantum cosmology\ \cite{Haro:2014wha}.

\begin{acknowledgments}
J. Q. acknowledges the Natural Sciences and Engineering Research Council (NSERC) of Canada
for financial support under a CGS M scholarship. The research of R. B. and Y. C. is supported
by an NSERC Discovery grant and by funds from the Canada Research Chair program.
\end{acknowledgments}

\appendix

\section{Solving for the Bogoliubov coefficients in the bounce phase}\label{sec:appendixA}

We want to solve the set of coupled first-order ordinary differential equations
\begin{eqnarray}
 \dot\alpha_k=\frac{\dot\omega_k}{2\omega_k}e^{2i\int_{t_i}^td\tilde t\,\omega_k(\tilde t)}\beta_k~, \\
 \dot\beta_k=\frac{\dot\omega_k}{2\omega_k}e^{-2i\int_{t_i}^td\tilde t\,\omega_k(\tilde t)}\alpha_k~.
\end{eqnarray}
To simplify the treatment, let us define
\begin{equation}
 g_k^{\pm}(t)\equiv\frac{\dot\omega_k}{2\omega_k}e^{\pm 2i\int_{t_i}^td\tilde t\,\omega_k(\tilde t)}~,
 \label{gk}
\end{equation}
together with the matrix
\begin{equation}
 A_k(t)\equiv\left( \begin{array}{cc}
                0 & g_k^+(t) \\
                g_k^-(t) & 0
               \end{array}\right)
\end{equation}
and the vector
\begin{equation}
 \mathbf{y}_k(t)\equiv\left( \begin{array}{c}
                          \alpha_k(t) \\
                          \beta_k(t)
                         \end{array}\right)~.
\end{equation}
Then, the system to solve simply reads
\begin{equation}
 \mathbf{\dot y}_k(t)=A_k(t)\mathbf{y}_k(t)~,
\end{equation}
and the initial conditions $\alpha_k(t_i)=1$ and $\beta_k(t_i)=0$ can be written as
\begin{equation}
 \mathbf{y}_k(t_i)=\left( \begin{array}{c}
                          1 \\
                          0
                         \end{array}\right)\equiv\mathbf{y}_0~.
\end{equation}
We make use of the Magnus expansion\ \cite{Magnus:1954zz}, which tells that the
solution to the above initial value problem is given by
\begin{equation}
 \mathbf{y}_k(t)=e^{\Omega_{k,1}(t)+\Omega_{k,2}(t)+...}\mathbf{y}_0~,
\end{equation}
where
\begin{align}
 \Omega_{k,1}(t)&=\int_{t_i}^t dt_1~A_k(t_1)~, \\
 \Omega_{k,2}(t)&=\frac{1}{2}\int_{t_i}^t dt_1\int_{t_i}^{t_1} dt_2~[A_k(t_1),A_k(t_2)]~,
\end{align}
and so on. Here, we consider the first-order approximation only. Defining
\begin{equation}
 I_k^{\pm}(t)\equiv\int_{t_i}^t dt_1~g_k^{\pm}(t_1)~,
 \label{Ik}
\end{equation}
the solution becomes
\begin{equation}
 \mathbf{y}_k(t)\simeq\exp\left( \begin{array}{cc}
                0 & I_k^+(t) \\
                I_k^-(t) & 0
               \end{array}\right)\mathbf{y}_0~,
\end{equation}
and therefore the Bogoliubov coefficients are given by
\begin{align}
 \alpha_k(t)&\simeq\cosh\sqrt{I_k^+(t)I_k^-(t)}~, \\
 \beta_k(t)&\simeq\sqrt{\frac{I_k^-(t)}{I_k^+(t)}}\sinh\sqrt{I_k^+(t)I_k^-(t)}~.
\end{align}
In the present case for $\omega_k(t)\simeq b+ct^2$, the function $g_k^{\pm}(t)$
defined in Eq.\ \eqref{gk} can be evaluated exactly.
Then, under the large-$\lambda$ assumption and as long as $b>0$,
the integral defined in Eq.\ \eqref{Ik} evaluated at the end of the particle production
phase becomes

\begin{equation}
 I_k^{\pm}(t_f)\simeq\mp\frac{ie^{\pm it_f\sqrt{b+ct_f^2}}}{2\sqrt{c}}\left[ct_f^2-b\ln\left(1+\frac{ct_f^2}{b}\right)\right]~.
\end{equation}
Therefore, the number of particles that has been produced is found to be
\begin{equation}
 n_k(t_f)=|\beta_k(t_f)|^2\simeq\sinh^2\frac{1}{2\sqrt{c}}\left[ct_f^2-b\ln\left(1+\frac{ct_f^2}{b}\right)\right]~.
\end{equation}

\end{document}